\documentclass[10pt,conference]{IEEEtran}
\usepackage{cite}
\usepackage{url}
\usepackage{amsmath,amssymb,amsfonts}
\usepackage{algorithm}
\usepackage{algorithmic}
\usepackage{graphicx}
\usepackage{booktabs}
\usepackage{placeins}
\usepackage{caption}

\def\BibTeX{{\rm B\kern-.05em{\sc i\kern-.025em b}\kern-.08em
    T\kern-.1667em\lower.7ex\hbox{E}\kern-.125emX}}

\newcommand{\tool}{\textsc{ExploreAI}}

\begin{document}

\title{ExploreAI: Agentic Exploration Knowledge Bases for Reproducible Observable-Regression Testing of Black-Box VR and 3D Applications}

\author{\IEEEauthorblockN{
Jiajie Wang\IEEEauthorrefmark{1},
Kebin Peng\IEEEauthorrefmark{2},
Wei Wang\IEEEauthorrefmark{3},
Xiaoyin Wang\IEEEauthorrefmark{3},
Sen He\IEEEauthorrefmark{1}, and
Xue Qin\IEEEauthorrefmark{4}}
\IEEEauthorblockA{
\IEEEauthorrefmark{1}The University of Arizona \\
\IEEEauthorrefmark{2}East Carolina University \\
\IEEEauthorrefmark{3}The University of Texas at San Antonio \\
\IEEEauthorrefmark{4}Villanova University}}

\maketitle

\noindent\begin{abstract}
Black-box VR and 3D applications are difficult to regression test because observable failures depend on where a tester moves, what objects are visible, and which views are captured. Manual exploratory testing can find such failures, but its evidence is time-consuming to reproduce; systematic sweeps are reproducible, but they lack semantic guidance and spend exploration budget on low-value viewpoints. We observe that an LLM can make the high-level decisions a human tester makes during exploration: interpreting a task, choosing which objects to inspect, grouping related objects, recording what it saw, and deciding when missing evidence should trigger another attempt. Based on this observation, we present \tool{}, an LLM-driven agentic framework that offloads repeated perception, navigation, multi-view capture execution, and logging to specialized modules while using the LLM for planning, evidence recording, capture-policy decisions, and verification decisions. \tool{} constructs an Exploration Knowledge Base (EKB): a structured, per-object record of one exploration run. For each object the agent finds, the EKB stores the scan evidence that exposed it, the selected target, the navigation path, the multi-view capture, and the self-verification result. The EKB is a reusable testing artifact that supports reproducible observable-regression checking across versions of a VR or 3D application. Across six indoor and outdoor scenes in Unity, AI2-THOR, and BeamNG, \tool{} constructs high-completeness EKBs under both complete and target exploration, and an LLM-module ablation shows where semantic planning, capture policy, evidence recording, and self-verification contribute. Reproduction pilots further show that EKB-guided traces help both humans and LLM-based reproducers reproduce exact object-view evidence more effectively than conditions without EKB context.
\end{abstract}

\noindent 

\section{Introduction}

\noindent Virtual reality (VR) and 3D applications are increasingly used in education, healthcare, manufacturing, training, and entertainment~\cite{fortuneVRMarket,cipresso2018vrar}. Testing these applications is difficult because correctness is not observed from a single screen state. A tester must move through a spatial environment, notice visible objects, approach them from suitable distances, inspect them from multiple views, and remember whether an object, path, or interaction behaved differently from a previous version. Manual exploratory testing can uncover such problems, but its evidence is often weak: two testers may take different paths, inspect different objects, and produce screenshots that are hard to replay or compare. Prior record-and-replay, model-based GUI testing, and visual regression work has shown that reproducible testing evidence requires more than ad hoc screenshots: the execution context, interaction trace, visual state, and oracle all affect whether a failure can be replayed and judged~\cite{gomez2013reran,memon2007eventflow,mahajan2014html,barr2015oracle}. In black-box 3D scenes, this problem is harder because the evidence also depends on navigation path, viewing distance, object side, and occlusion. A fixed systematic sweep is replayable, but pure random exploration is not inherently reproducible unless the seed and full trace are preserved; both strategies also lack the semantic judgment a human tester uses when deciding which object is worth approaching, which target has already been inspected, and which missed view should trigger recovery.

Existing VR and XR testing systems rely on environment understanding to execute interaction scenarios and improve testing effectiveness~\cite{zhu2025vrexplorer,gu2026xrinttest,longfils2026ultrainstinctvr,coppola2024gamemetrics}. Meanwhile, exploration and embodied-AI research develops increasingly effective techniques for navigating unknown environments and constructing semantic representations~\cite{yamauchi1997frontier,vlnsurvey,hm3d,ai2thor,scenegraphcontrastive,sceneMMKG}. However, these efforts optimize testing, navigation, or semantic reasoning rather than preserving the evidence discovered during exploration.
Although GUI testing operates in a fundamentally different interaction context, it demonstrates the value of reusable testing artifacts through event-flow models, interaction traces, and visual snapshots for regression testing~\cite{memon2007eventflow,gomez2013reran,mahajan2014html}. These ideas inspire our use of visual observations and interaction histories during exploration.
Consequently, \textbf{current VR exploration techniques provide little support for preserving exploration results as reusable testing artifacts.} The discovered environment information is largely ephemeral, making it difficult to replay, compare, and reuse across testing tasks or software versions.

\begin{figure}[t]
\centering
\includegraphics[width=0.95\columnwidth]{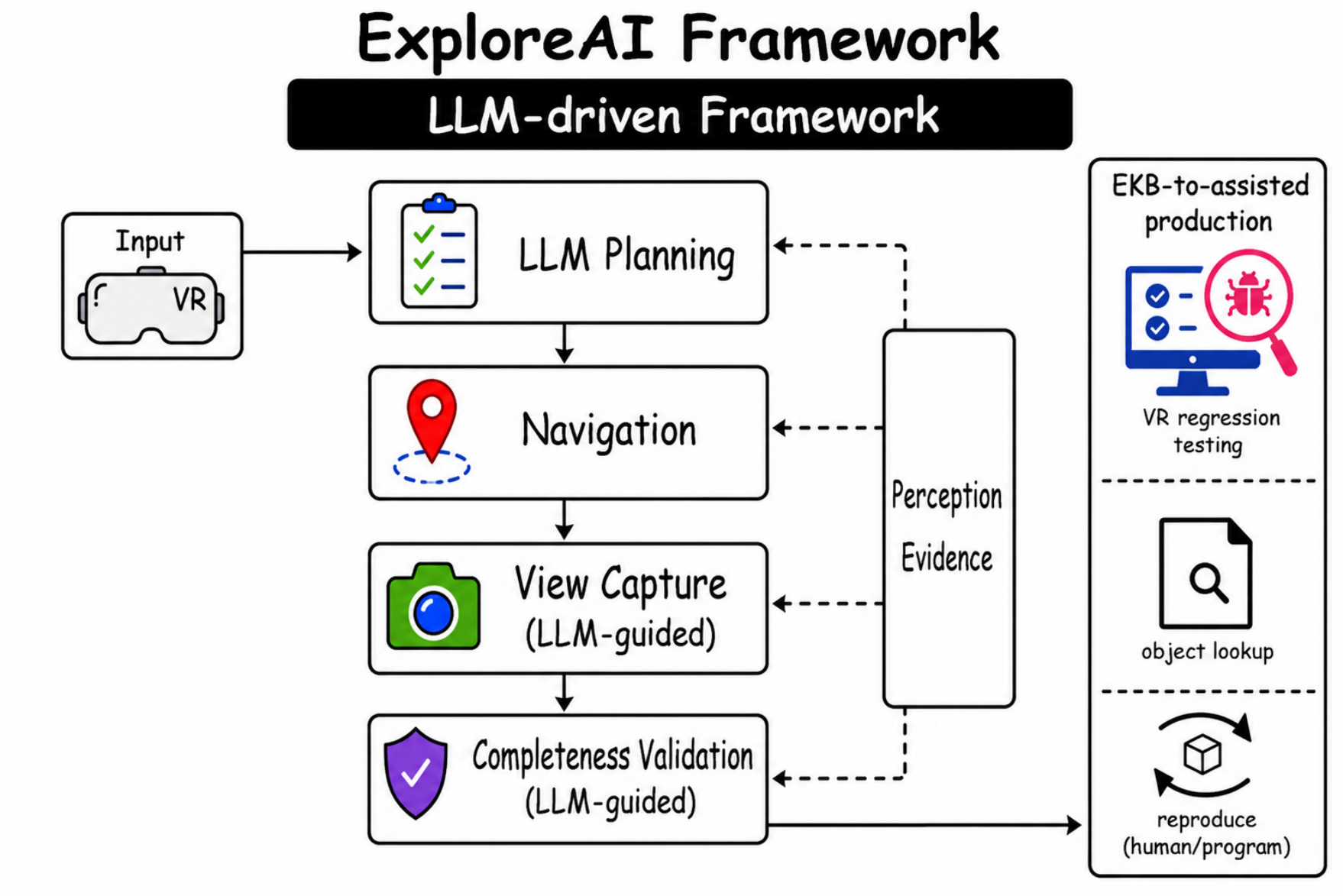}
\caption{\tool{} as an LLM-driven framework for reproducible black-box VR/3D testing. A running application and testing task enter the framework; the LLM guides planning, view-capture policy, and completeness validation while perception evidence, navigation context, and multi-view observations are recorded during execution. The output is an Exploration Knowledge Base (EKB) that supports downstream tasks including VR/3D regression testing, object-trace inspection, and human or programmatic reproduction.}
\label{fig:intro_overview}
\end{figure}

To address this challenge, we present \tool{}, an agentic framework that constructs an Exploration Knowledge Base (EKB) from a running black-box VR or 3D application. The EKB is a structured, per-object record: for each object it stores where the object was observed, how it was reached, the views captured, and whether any evidence was missing (Section~\ref{sec:problem}), which makes observable object-view evidence reproducible. Fig.~\ref{fig:intro_overview} summarizes the idea: a running VR application and testing task enter the framework, the LLM guides high-level planning, view-capture policy, and completeness validation, and specialized perception and navigation modules collect evidence for the final EKB. \tool{} begins with an autonomous seed sweep, iteratively selects targets, centers and approaches them, captures multi-view evidence, returns along the capture path, scans again, and records the resulting object trace. The LLM is not a continuous low-level controller; perception, motion, view capture, and logging are delegated to specialized modules so that long runs remain practical.

We evaluate \tool{} on six indoor and outdoor scenes across Unity, AI2-THOR, and BeamNG. Under equal scene budgets, \tool{} reaches 0.979 object completeness and 0.971 view completeness in complete exploration (object completeness is the fraction of the objects a tester should observe that receive a full evidence trace; view completeness is the fraction of the required camera angles that are captured, both measured against a frozen manifest), completes all query-relevant targets in target exploration, and keeps the per-run LLM cost modest (RQ1). The LLM-module ablation shows that removing LLM evidence recording lowers navigation-path and log completeness, while removing LLM planning lowers Urban Cabin object/view completeness from 0.871/0.825 to 0.323/0.209 and removing self-verification lowers its aggregate complete-exploration score to 0.666 (RQ2). For reproducible regression testing, EKB context helps humans and automated agents reproduce object-view evidence more completely and faster than conditions without it, including scene versions with changed elements (RQ3).

We summarize our contributions as follows:
\begin{itemize}
\item We frame observable-regression testing of black-box VR and 3D applications as a reproducibility problem and propose the Exploration Knowledge Base (EKB), an object-centric testing artifact that preserves traces, multi-view observations, navigation paths, scan evidence, recovery outcomes, failure cases, and model logs for replay and comparison across versions.
\item We propose \tool{}, an automated EKB-construction framework that combines LLM-driven test planning, target interpretation, capture planning, and completeness checking with YOLO-backed perception for navigation and deterministic motion, capture, and logging modules.
\item We evaluate \tool{} across six scenes and six LLM configurations, ablate the LLM-guided modules, evaluate cross-version regression detection with DiffEKBs, and show that the resulting EKBs enhance human and automated reproduction of exploratory testing evidence.
\end{itemize}

\section{Related Work and Problem Statement}

\subsection{Related Work}

\noindent \textbf{VR Testing and Interaction.}
Existing work studies model-based scene exploration, interaction execution, and scenario generation to improve VR testing effectiveness. VRExplorer constructs abstract scene models to explore VR environments and interact with virtual objects~\cite{zhu2025vrexplorer}. XRintTest models XR user interactions and dynamically explores scenes to execute representative 3D interactions~\cite{gu2026xrinttest}. UltraInstinctVR generates VR system tests from scenario models~\cite{longfils2026ultrainstinctvr}, while recent work investigates whether vision-capable LLMs can assist exploratory VR testing by analyzing first-person observations~\cite{qi2025vrtesting}. Broader surveys of LLM-assisted software testing document both this promise and the nondeterminism and cost risks of language models in testing~\cite{wang2024llmtesting}.
These approaches demonstrate the importance of environment understanding for automated VR testing. However, they use environment understanding to drive a single test execution; none preserves the object-view evidence of that execution as a reusable artifact for cross-version regression comparison.

\noindent\textbf{Automated Exploration in Virtual Environments.}
Automated exploration has been extensively studied in robotics, embodied AI, and virtual environments. Classical exploration algorithms maximize environment coverage through frontier-based or coverage-driven strategies~\cite{yamauchi1997frontier,lavalle2001rrt,burgard2005coordinated,stachniss2005info,hollinger2014sampling}.
More recent embodied AI research investigates vision-language navigation, object-goal navigation, and retrieval-augmented navigation to guide agents toward task-specific destinations~\cite{vlnsurvey,navrag}. 
Although these approaches effectively traverse unknown environments, their primary objective is to maximize exploration efficiency through navigation or coverage optimization. As a result, evaluation typically focuses on metrics such as explored area, path efficiency, or coverage, rather than the completeness, organization, or long-term usability of the discovered environment knowledge. Crucially, none preserves the discovered evidence as a reusable artifact for reproducible regression testing.

\noindent\textbf{Environment Knowledge Construction.}
Large-scale virtual environment datasets and embodied AI platforms provide scene annotations, reconstructed environments, navigation trajectories, and agent interfaces for studying perception and navigation~\cite{hm3d,habitat,ai2thor,robothor}. Building on these resources, researchers have constructed scene graphs, multimodal scene knowledge graphs, and retrieval-based environment models to support navigation and semantic reasoning~\cite{scenegraphcontrastive,sceneMMKG}. While these efforts demonstrate the value of structured environment knowledge, they rely on curated datasets or offline annotations. They do not address how such knowledge can be acquired automatically from an arbitrary running VR application, nor how to preserve it as reproducible evidence for regression testing.

\noindent\textbf{GUI Exploration and Testing.}
Although GUI testing techniques cannot be directly applied to immersive 3D environments due to fundamentally different interaction contexts, they provide valuable insights into constructing reusable testing artifacts. Model-based techniques build event-flow graphs, state-transition models, or stochastic abstractions for systematic exploration and test generation~\cite{memon2007eventflow,mesbah2012crawling,mobiguitar2015,stoat2017}, while record-and-replay systems preserve interaction traces for later execution~\cite{gomez2013reran}. Visual GUI testing further uses screenshots to locate interface elements, automate interactions, and detect visual regressions~\cite{sikuli2009,chang2010visiongui,mahajan2014html}, closely relating to the test-oracle problem of interpreting observed differences~\cite{barr2015oracle}. Together, these techniques demonstrate the value of preserving visual observations and interaction histories as reusable testing artifacts, motivating our construction of an Exploration Knowledge Base for immersive 3D environments.

\subsection{Problem Statement}
\label{sec:problem}

\noindent\textbf{Setting.} We test a black-box VR or 3D application. The tester can render frames and send input controls, but cannot read the engine scene graph, the source code, or any internal state. The only evidence is what a user-level tester can see on screen while moving through the scene. Correctness is therefore judged from observable object-view evidence, not from code or memory.

\noindent\textbf{Input and output.} The input is a running build and a testing goal. The goal is either \emph{complete exploration}, which documents every observable object in the scene, or \emph{target exploration}, which documents one object category. The output is an Exploration Knowledge Base (EKB).

\noindent\textbf{The EKB and its terms.} An EKB is a structured, per-object record of one exploration run. For each object the agent finds, it stores five things. \emph{Scan evidence} is the frames and detections from a stationary sweep, taken before the agent moves. A \emph{selected target} is an object the planner decides is worth approaching, together with the reason it was chosen. The \emph{navigation path} is the route the agent took to reach the object and return. A \emph{multi-view capture} is the set of angles saved around the object: four for a free-standing object, two for an object against a wall. \emph{Self-verification} is a second pass that looks for a missing object or a missing angle and tries to recover it before the EKB is saved. The EKB also stores scene metadata, the model logs, and the failure cases. Section~\ref{sec:method} gives the full schema.

\noindent\textbf{The problem.} Two exploration runs of the same build take different paths and save different angles, so their evidence is hard to compare, and a later version is hard to check against an earlier one. We call a version difference an \emph{observable regression} when it changes what a tester can see: a removed object, a swapped asset, a blocked path, or a lost view. The problem we address is to construct, from a running black-box application, a reusable artifact from which the same object-view evidence can be reproduced by a human or an agent, and against which a later version can be diffed to flag observable regressions. We measure the artifact by \emph{object completeness} and \emph{view completeness} against a frozen manifest: a human-checked list of the objects that should be observable and the views that should be reachable in each scene. Section~\ref{sec:method} defines both metrics.

\section{Methodology}
\label{sec:method}

\subsection{Overview}

\begin{figure*}[t]
    \centering
    \includegraphics[width=0.95\textwidth]{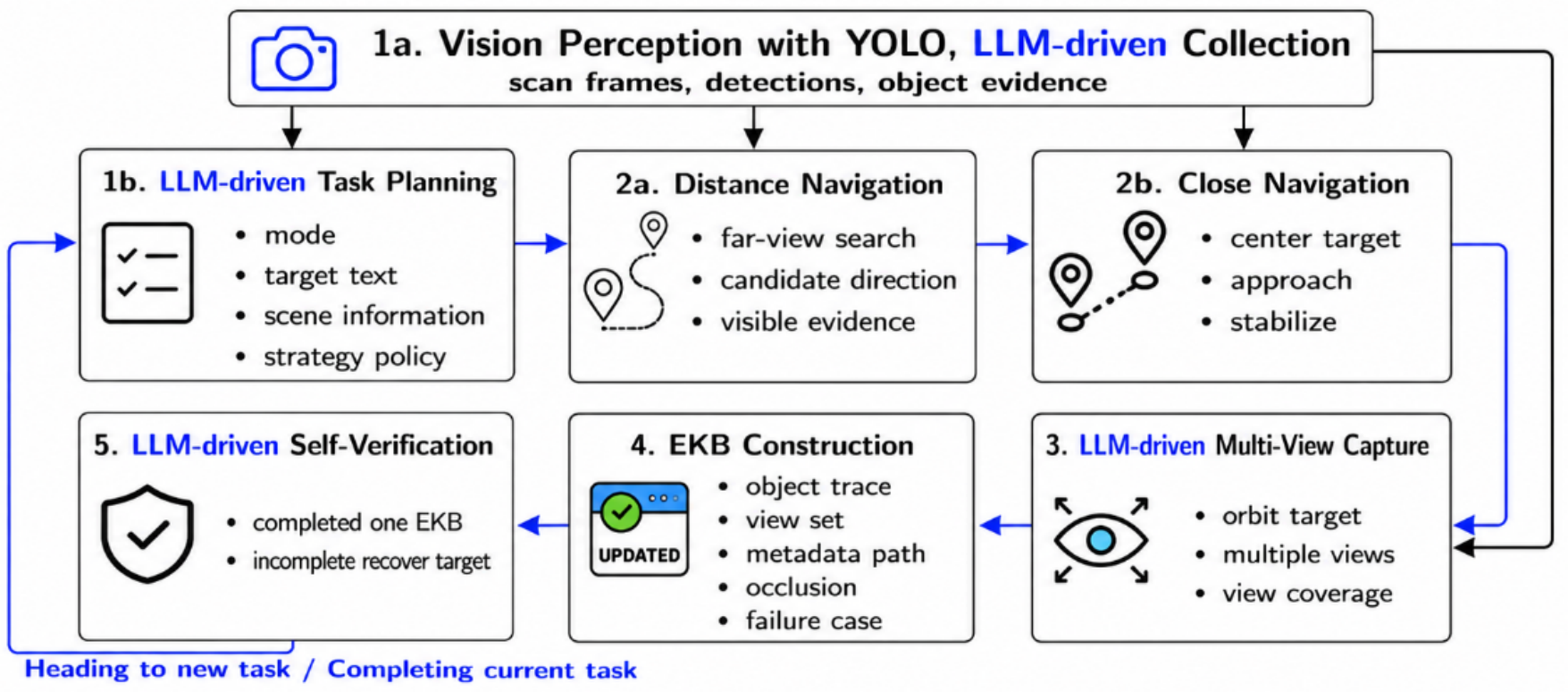}
    \caption{Detailed \tool{} workflow. Perception provides scan frames, detections, and object evidence to task planning, distance navigation, close navigation, LLM-driven multi-view capture, EKB update, and self-verification. LLM-driven planning, capture policy, evidence recording, and self-verification operate at decision points, while YOLO-backed perception and deterministic control handle repeated low-level execution.}
    \label{fig:framework}
\end{figure*}

\begin{figure*}[t]
    \centering
    \includegraphics[width=0.95\textwidth]{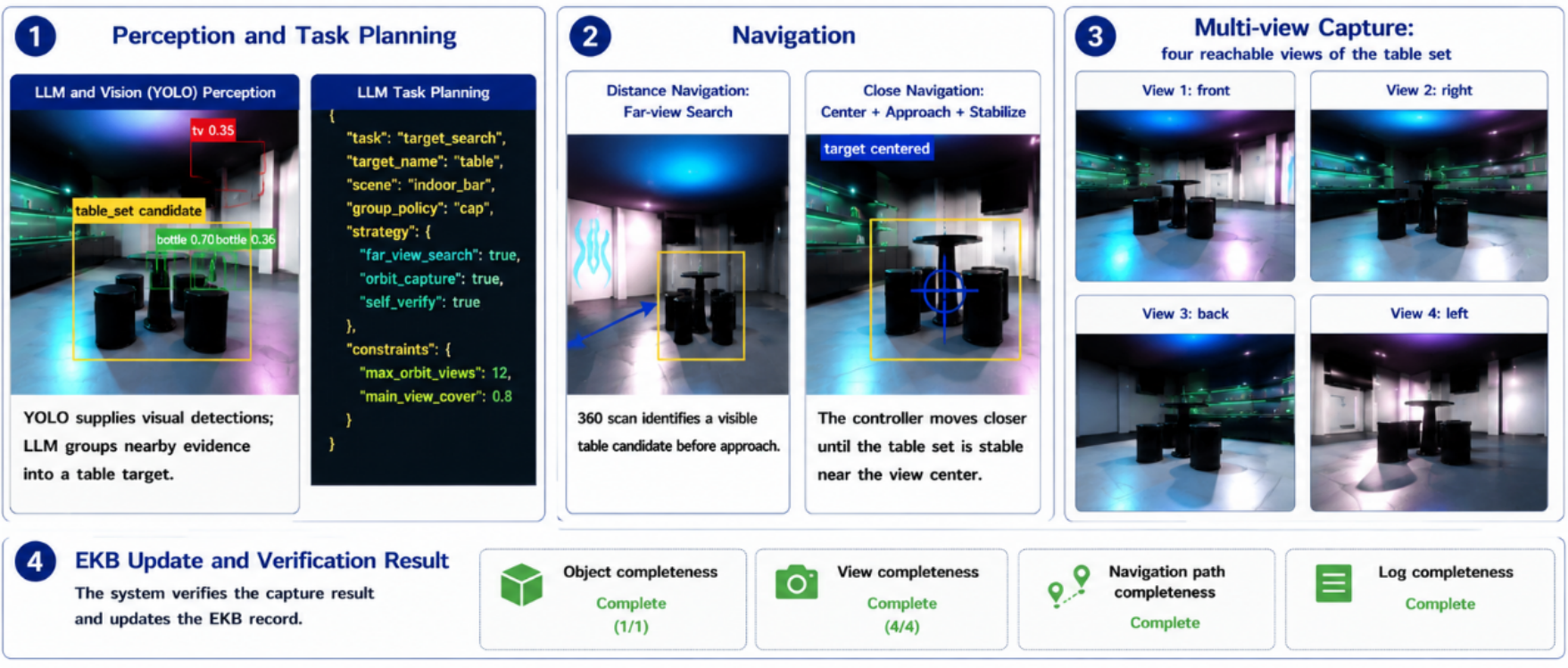}
    \caption{Example real execution evidence from a Unity indoor bar run. The example shows the task plan and YOLO detections, distance and close navigation toward a bottle shelf, LLM-driven multi-view captures around the target, and the resulting EKB update and self-verification record.}
    \label{fig:execution_example}
\end{figure*}

\noindent\tool{} converts a task specification and a running black-box VR/3D application into a reproducible Exploration Knowledge Base (EKB). Fig.~\ref{fig:framework} summarizes the workflow, and Fig.~\ref{fig:execution_example} shows a concrete indoor-bar execution example. A run starts from a user-level testing goal, either complete exploration or target exploration for a category. The system performs an autonomous seed sweep, detects candidate objects from rendered frames, asks the LLM planner to interpret the task and prioritize targets, navigates toward selected objects, captures multi-view evidence, returns to the scan state, and finally checks whether missing or incomplete traces should be recovered. The output is an EKB rather than a folder of screenshots: each object trace links scan evidence, target selection, navigation path, approach context, saved views, recovery status, and failure notes.

\tool{} supports two testing modes because VR testing needs both broad discovery and focused regression checks. In \emph{complete exploration}, the user gives a broad instruction (for example, ``explore this bar scene'') and the framework attempts to identify, approach, and capture evidence for all observable targets in the scene budget. This mode is useful when a developer first builds an EKB for an unfamiliar scene or wants to maximize object-view coverage. In \emph{target exploration}, the user gives a semantic target (for example, ``inspect cars'' or ``check the shelf bottles''). The LLM maps the text to detectable categories or functional groups, and the controller spends the budget on those targets rather than recording every object it passes. This mode is useful for regression testing when a developer knows that a changed asset class, interaction area, or object group needs to be rechecked.

\begin{algorithm}[t]
\caption{ConstructEKB}
\label{alg:construct_ekb}
\scriptsize
\begin{algorithmic}[1]
\REQUIRE running application \(A\), task \(q\), budget \(B\), manifest policy \(P\)
\ENSURE exploration knowledge base \(K\)
\STATE \(K \leftarrow\) initialize metadata, task mode, detector, LLM plan, and budget
\STATE \(S \leftarrow\) autonomous seed sweep of \(A\); store frames and detections in \(K\)
\STATE \(C \leftarrow\) candidate objects from perception over \(S\)
\WHILE{\(B\) remains and unresolved candidates exist}
  \STATE \(o \leftarrow\) rank candidate using semantic relevance, visibility, distance, and repeat penalty
  \IF{\(o\) is invalid, already complete, or outside the task scope}
    \STATE record skip reason in \(K\); continue
  \ENDIF
  \STATE \(N \leftarrow\) navigate to \(o\) with far-view and near-view control
  \IF{navigation fails or detector support is lost}
    \STATE append failed trace \(\langle o,N,F\rangle\) to \(K\); continue
  \ENDIF
  \STATE \(V \leftarrow\) capture reachable multi-view evidence around \(o\)
  \STATE reverse the capture path and return to scan state
  \STATE append object trace \(\langle id,c,\hat{p},D,S,N,V,R,F\rangle\) to \(K\)
  \STATE \(S \leftarrow\) rescan scene; update candidates and unresolved set
\ENDWHILE
\STATE \(K \leftarrow\) RecoverMissingTrace(\(K,A,P,B\))
\RETURN \(K\)
\end{algorithmic}
\end{algorithm}

\subsection{Perception and LLM-Driven Task Planning}

\noindent\textbf{Perception.} \tool{} uses YOLO-style object detection~\cite{redmon2016yolo} as the repeated perception layer over rendered frames. A vision-capable LLM can perform pure perception by reading screenshots and describing visible objects, but prior VR exploration work reports that calling an LLM for every field-of-view judgment, movement step, or low-level visual confirmation is slow and costly~\cite{wang2026navai,qi2025vrtesting}. We therefore use YOLO for the high-frequency visual loop and reserve the LLM for semantic decisions and evidence recording. This design follows the same practical motivation as prior VR exploration-testing work: perception is needed throughout the entire cycle, so it must be cheap enough to run at scan, navigation, capture, and recovery time~\cite{qi2025vrtesting}.

\noindent\textbf{LLM evidence recording.} The LLM records the semantic layer that a detector alone does not provide. For each run, it records the task mode, target text, scene-level assumptions, grouping policy, expected evidence, and risk notes. For each selected target, it records why the target is relevant, which visual evidence supports it, whether it belongs to a functional group, what views are expected, and why a target is skipped, retried, or marked incomplete. These records become part of the EKB rather than transient prompts. YOLO supplies fast visual evidence; the LLM turns that evidence into a reproducible testing trace.

\noindent\textbf{LLM-driven task planning.} The LLM planner performs scene-level and task-level interpretation. Given a complete-exploration prompt or a target query, it determines the task mode, maps text (for example, \emph{car}, \emph{chair}, or \emph{table set}) to detectable categories, decides which scene risks should be tracked, and records the expected evaluation outputs. It also supports scene understanding by interpreting object lists exposed by perception, identifying semantically relevant object groups, and deciding whether nearby instances should be treated as a functional unit. For example, a dining table and surrounding chairs can be grouped as one table-set trace in grouped mode, while the same objects remain separate traces in ungrouped mode.

Planning and decision rules combine semantic relevance, visual evidence, spatial diversity, and completion status. For a candidate object \(o\), the planner uses the score
\[
\begin{aligned}
score(o)=&\,w_s sem(o)+w_v vis(o)+w_p prom(o)\\
&-w_d dist(o)-w_r repeat(o),
\end{aligned}
\]
where \(sem(o)\) is task relevance, \(vis(o)\) is detector support across scan views, \(prom(o)\) is image prominence, \(dist(o)\) is navigation distance, and \(repeat(o)\) penalizes already completed or recently failed traces. The weights are fixed in the run configuration for each platform. In complete exploration, semantic relevance is broad and the diversity term prevents the system from repeatedly selecting the same category. In target exploration, the LLM maps the text query to detectable categories and the controller ranks matching candidates by visibility and reachability. Grouping is LLM-interpreted and geometry-instantiated: once the LLM decides that a category should be grouped, such as a shelf set or table set, the controller applies distance and bounding-box rules to assign member objects while preserving their identifiers in the trace. The run configuration is reported in Table~\ref{tab:config}.

\subsection{Navigation: Far-View and Near-View Control}

\noindent After target selection, the deterministic controller navigates in two stages. In \emph{far-view navigation}, the system moves toward the candidate region while keeping the target direction stable. YOLO-backed perception is used here because it provides fast object confirmation from ordinary rendered frames and does not require simulator-specific object APIs or an LLM call for every step. The far-view stage stops when the camera reaches the configured approach radius, the target becomes visually centered enough for close control, or the movement budget is exhausted.

In \emph{near-view navigation}, the system centers the target and stabilizes the camera before capture. The controller uses the detector bounding box to estimate the target's screen-center error and adjusts the view until the box center falls within a configured tolerance. It also checks whether the target remains within the allowed distance band for orbiting. If the target disappears, becomes blocked, or cannot be centered within the step budget, the trace is marked as a path failure or target-lost failure instead of being silently dropped. This two-stage design keeps long-distance movement separate from close visual stabilization, which makes the resulting navigation path easier to reproduce and inspect.

\subsection{LLM-Driven Multi-View Capture}

\noindent The capture module saves object-centric evidence after the target is centered and stabilized. The LLM-driven capture policy determines what views should be considered sufficient evidence for the target: four views for free-standing objects, fewer reachable views for objects attached to walls or blocked by scene boundaries, and grouped views when a functional object set should be documented together. For free-standing objects, the expected denominator is four views: front, left, back, and right. For objects close to walls or boundaries, the reachable-view policy reduces the denominator to the views that can be captured without crossing the boundary, usually two distinct angles. Each saved view stores the target id, semantic label, view angle, timestamp, detector confirmation, crop path, and full-frame path.

The controller captures views while following an orbit path around the target, but the LLM records the capture intent and expected evidence. It labels the requested views, decides whether a wall-adjacent target should require two views rather than four, marks which view is missing when a path is blocked, and decides whether an alternate orbit direction should be attempted. If one direction is blocked, the capture policy can switch direction and attempt the missing side from the opposite route. After capture, the controller reverses along the recorded orbit path before returning to scan. This reverse replay preserves the visual continuity of the run and avoids jumps between unrelated targets. A capture attempt is marked complete when all expected reachable views are saved and detector-confirmed; otherwise, the missing views are left in the trace for self-verification.

\subsection{EKB Schema, Coverage, and LLM-Driven Self-Verification}

\noindent \textbf{EKB Schema and Coverage.} The central unit in the EKB is an \emph{object trace}. An object trace records where the object was first observed, which detections supported it, which target category was assigned, how it was approached, which views were captured, and whether self-verification recovered or failed to recover missing evidence. We define an EKB as a set of scene-level metadata, scan records, object traces, reference views, and evaluation records. Each object trace is a tuple
\[
T=\langle id,c,\hat{p},D,S,N,V,R,F\rangle ,
\]
where \(id\) is a stable object identifier within the run, \(c\) is the semantic category, \(\hat{p}\) is the estimated 3D position or region, \(D\) is the set of supporting visual detections, \(S\) is the scan evidence that exposed the target, \(N\) is the navigation and approach path, \(V\) is the set of saved object-centric views, \(R\) is the recovery status, and \(F\) is the failure record if the target cannot be completed. The EKB also stores run configuration, task mode, LLM plan, detector model, videos, and per-target metrics. The EKB is scored on four completeness components: object, view, navigation-path, and log completeness.

Completeness is measured against the frozen manifest, not the system's own plan. Given a manifest target set \(M\), object completeness is \(|\{m\in M: completed(m)\}|/|M|\). Given expected reachable views \(E_m\) for target \(m\), view completeness is \(\sum_m |V_m \cap E_m|/\sum_m |E_m|\). A target is completed when the trace contains the required reachable views, or the wall policy marks the target as requiring only the reachable subset. A failed trace is still part of the EKB; it contributes evidence for incompleteness rather than disappearing from the denominator.

\noindent\textbf{LLM-driven self-verification.} The first exploration pass can miss objects or views because targets are occluded, close to walls, partially outside the field of view, or detected only from a weak angle. Instead of treating the first pass as final, \tool{} performs LLM-driven self-verification before saving the EKB. Self-verification is triggered when a planned target is skipped, when an expected reachable view is missing, or when detector confirmation is absent for all captured views. The LLM does not re-drive the low-level controller; it reviews the incomplete trace, the scan evidence, the wall policy, and the failure reason, and then decides whether recovery is worth attempting and which alternate side or target description to use. Successful recovery is merged into the object trace; unresolved cases are kept as explicit failures. This step improves coverage by giving the system a structured chance to recover missing object traces and missing reachable views before finalizing the EKB. The final EKB contains scan screenshots, detections, selected targets, orbit captures, reference captures, navigation paths, full videos, per-target results, failure cases, quantitative summaries, and LLM logs.

\begin{algorithm}[t]
\caption{RecoverMissingTrace}
\label{alg:recover_missing}
\scriptsize
\begin{algorithmic}[1]
\REQUIRE EKB \(K\), application \(A\), manifest policy \(P\), remaining budget \(B\)
\ENSURE updated EKB \(K\)
\STATE \(U \leftarrow\) traces with missing object, missing reachable view, target lost, or no detector confirmation
\FOR{each trace \(t \in U\)}
  \IF{\(B\) is exhausted}
    \STATE mark \(t\) unresolved with budget failure; continue
  \ENDIF
  \STATE infer alternate search side from \(t\)'s scan evidence, wall policy, and failed path
  \STATE rescan from the alternate side and re-rank matching candidates
  \IF{no matching candidate is visible}
    \STATE keep \(t\) as explicit missing evidence; continue
  \ENDIF
  \STATE approach the candidate and capture only missing reachable views
  \IF{new views satisfy \(P\)}
    \STATE merge recovery views and mark \(t\) recovered
  \ELSE
    \STATE preserve partial views and failure reason
  \ENDIF
\ENDFOR
\RETURN \(K\)
\end{algorithmic}
\end{algorithm}

\begin{algorithm}[t]
\caption{DiffEKBs}
\label{alg:diff_ekbs}
\scriptsize
\begin{algorithmic}[1]
\REQUIRE baseline EKB \(K_b\), candidate EKB \(K_c\), tolerance policy \(P\)
\ENSURE regression report \(R\)
\STATE match traces by stable id when available; otherwise match by category, region, and supporting views
\FOR{each expected trace \(t_b \in K_b\)}
  \STATE \(t_c \leftarrow\) best matching trace in \(K_c\)
  \IF{\(t_c\) is missing}
    \STATE add missing-object regression to \(R\); continue
  \ENDIF
  \IF{category or asset-level label changes beyond \(P\)}
    \STATE add wrong-object regression to \(R\)
  \ENDIF
  \IF{position shift exceeds tolerance or path status becomes blocked}
    \STATE add movement or reachability regression to \(R\)
  \ENDIF
  \IF{expected reachable views in \(t_b\) are absent or incomplete in \(t_c\)}
    \STATE add lost-view regression to \(R\)
  \ENDIF
\ENDFOR
\STATE ignore benign changes that preserve category, reachability, and expected view set
\RETURN \(R\)
\end{algorithmic}
\end{algorithm}

\section{Experiment Setups}

We evaluate \tool{} as a testing framework rather than only as an exploration pipeline. The evaluation asks whether the EKB improves observable black-box VR/3D regression testing: can it cover the scene more completely and at practical cost, which LLM-guided modules contribute most, and can the artifact make regression evidence more reproducible for both automated approaches and humans? Our comparisons are designed to isolate the artifact. Existing VR testing systems such as VRExplorer~\cite{zhu2025vrexplorer} and XRintTest~\cite{gu2026xrinttest} execute tests but do not produce the reusable object-view evidence artifact that \tool{} targets. We therefore compare the same framework with and without EKB context (RQ3) and ablate the LLM hooks (RQ2), which isolates the value of the artifact and of each decision.

\subsection{Research Questions}

\noindent\textbf{RQ1. How complete are the EKBs that \tool{} constructs, and what is the cost?}
We evaluate RQ1 with EKB completeness metrics for complete exploration and target exploration: object completeness, view completeness, navigation-path completeness, and log completeness. We then evaluate approach generality with a six-model benchmark over the same six scenes, where each cell reports an aggregate EKB completeness score. Finally, we report token and API cost for complete and target exploration under official provider prices.

\noindent\textbf{RQ2. How much do the individual modules, especially LLM-driven decisions, contribute?}
We evaluate RQ2 with module ablations over six scenes. The ablation metrics are aligned with the EKB components in RQ1: object completeness, view completeness, navigation-path completeness, log completeness, and aggregate completeness for complete and target exploration.

\noindent\textbf{RQ3. Do EKBs support reproducible regression testing, both by reproducing evidence and by detecting cross-version regressions?}
We evaluate RQ3 with reproduction metrics for human and automated agents (object completeness, view completeness, time, and token use with and without EKB context) and, as a regression-detection check, by comparing baseline and post-change EKBs with Algorithm~\ref{alg:diff_ekbs} over injected element changes (removed or added objects, swapped assets, blocked paths, and lost views), reporting per-category detection precision and recall against the injected ground truth.

\subsection{Experimental Setup}

\noindent We evaluate six scenes: three indoor scenes (Unity Korean Bar, Unity Urban Furnished Cabin (Urban Cabin), and AI2-THOR FloorPlan203) and three outdoor scenes (Unity Road, Unity Cartoon Low Poly City (Cartoon City), and BeamNG). The scene set covers dense indoor furniture, room-scale navigation, road scenes, low-poly outdoor environments, and vehicle-oriented scenes. Each run follows the same complete-flow protocol: autonomous scan, target selection, centering, approach, multi-view capture, reverse motion, re-scan, and self-verification.

All local experiments were executed on a Windows desktop with an AMD Ryzen 7 7800X3D CPU (8 cores, 16 logical processors), 32 GB RAM, and an NVIDIA GeForce RTX 3080 GPU. The operating system was Windows 11 Home, 64-bit, build 26200. Unity experiments used Unity 6.0.0f1 (6000.0.40f1). AI2-THOR experiments used AI2-THOR 5.0.0 in a Python 3.10.11 virtual environment; BeamNG and Unity-side analysis scripts used Python 3.13.1. LLM calls were issued through a remote API service, so the local hardware affects rendering, perception, navigation, frame capture, and video encoding time, but not model inference throughput. The six configurations pair a large and a small model from three providers: GPT-5.4 and GPT-5 mini (\texttt{gpt-5.4}, \texttt{gpt-5-mini-2025-08-07}), Gemini 2.5 Pro and Gemini 3.1 Flash Lite (\texttt{gemini-2.5-pro}, \texttt{gemini-3.1-flash-lite}), and Claude Opus 4.8 and Claude Haiku 4.5 (\texttt{claude-opus-4-8}, \texttt{claude-haiku-4-5-20251001}). The identifiers in parentheses are the exact API strings used in our runs.

\begin{table}[t]
\centering
\caption{\tool{} representative run configuration. Score weights parameterize the normalized candidate $score(o)$.}
\label{tab:config}
\scriptsize
\setlength{\tabcolsep}{3pt}
\begin{tabular}{@{}p{0.43\columnwidth}p{0.50\columnwidth}@{}}
\toprule
Parameter & Value \\
\midrule
Score weights $w_s,w_v,w_p,w_d,w_r$ & 0.40, 0.25, 0.15, 0.10, 0.10 \\
YOLO min box area ratio & 0.0004 \\
Movement budget & no target cap; max scan cycles 30; speed 2.2 \\
Approach stop & target-area ratio 0.10 \\
Capture setting & 1280$\times$720 at 5 FPS \\
Orbit policy & 36 steps; 0.12 s/step; adaptive radius \\
View denominator & 4 free-standing / 2 wall-adjacent \\
Prompts and EKB schema & released in artifact \\
\bottomrule
\end{tabular}
\end{table}

\subsection{Ground Truth}

\noindent All completeness metrics are computed against a frozen manifest, not against \tool{}'s own completed-target list. For each scene, two annotator passes checked the expected observable objects and reachable views before metric computation, and disagreements were resolved before freezing the manifest. The manifest is therefore an evaluation oracle, not an output of the system under test. The final two-person confirmation of the frozen include/exclude manifest records Cohen's kappa~=~0.86~\cite{cohen1960kappa} and object-id Jaccard agreement~=~0.89 before adjudication. The final manifest contains 10 targets for AI2-THOR, 5 for BeamNG, 9 for Unity Cartoon Low Poly City, 7 for Unity Road, 63 for Unity Korean Bar, and 31 for Unity Urban Furnished Cabin. These manifests define the denominator for object completeness and expected view completeness. We preserve the manifest, raw disagreement notes, and adjudication records with the artifact so that the agreement statistics can be audited.

For construction comparisons, the budget is the scene-level exploration budget used by the complete \tool{} run, measured as wall-clock runtime and the corresponding number of scan/capture opportunities. Each model-scene pair uses ten total repetitions: five complete-exploration runs and five target-exploration runs. We keep this budget fixed across complete exploration, target exploration, and LLM-module ablations so that object, view, path, log, token, and cost measurements remain comparable.

\section{Evaluation}

\subsection{RQ1: How complete are the EKBs that \tool{} constructs, and what is the cost?}

\noindent RQ1 first asks whether \tool{} constructs complete EKBs in both complete exploration and target exploration. Table~\ref{tab:rq1_dynamic} reports detailed completeness for a strong representative configuration, GPT-5.4 (one of the six in Table~\ref{tab:rq1_llm_components}). Object completeness measures completed object traces over the frozen scene manifest in complete exploration and over the query-relevant target subset in target exploration. View completeness measures saved reachable views over expected reachable views. Navigation-path completeness measures whether completed traces preserve approach, orbit, and return-path evidence. Log completeness measures whether completed traces preserve scan evidence, detections, verification decisions, and metadata. Across the six scenes, complete exploration reaches 0.979 object completeness and 0.971 view completeness as macro averages, while target exploration completes all query-relevant targets.

\begin{table*}[t]
\centering
\caption{RQ1 detailed EKB completeness for a strong representative LLM configuration, GPT-5.4. Complete exploration uses frozen scene manifests; target exploration uses query-relevant target subsets. Pat/log measured over completed traces.}
\label{tab:rq1_dynamic}
\scriptsize
\setlength{\tabcolsep}{4pt}
\resizebox{0.65\textwidth}{!}{%
\begin{tabular}{l|rrrr|rrrr}
\toprule
Scene & \multicolumn{4}{c|}{Complete exploration} & \multicolumn{4}{c}{Target exploration} \\
\cmidrule(lr){2-5}\cmidrule(lr){6-9}
 & Object & View & Path & Log & Object & View & Path & Log \\
\midrule
AI2-THOR & 1.000 & 1.000 & 1.000 & 1.000 & 1.000 & 1.000 & 1.000 & 1.000 \\
BeamNG & 1.000 & 1.000 & 1.000 & 1.000 & 1.000 & 1.000 & 1.000 & 1.000 \\
Cartoon City & 1.000 & 1.000 & 1.000 & 1.000 & 1.000 & 1.000 & 1.000 & 1.000 \\
Unity Road & 1.000 & 1.000 & 1.000 & 1.000 & 1.000 & 1.000 & 1.000 & 1.000 \\
Korean Bar & 1.000 & 1.000 & 1.000 & 1.000 & 1.000 & 1.000 & 1.000 & 1.000 \\
Urban Cabin & 0.871 & 0.825 & 1.000 & 1.000 & 1.000 & 1.000 & 1.000 & 1.000 \\
\midrule
Overall & 0.979 & 0.971 & 1.000 & 1.000 & 1.000 & 1.000 & 1.000 & 1.000 \\
\bottomrule
\end{tabular}
}
\end{table*}

Table~\ref{tab:rq1_llm_components} evaluates approach generality across six LLM configurations, six scenes, and both exploration modes. Each cell reports aggregate EKB completeness, computed as the arithmetic mean of the same four components reported in Table~\ref{tab:rq1_dynamic}: object, view, navigation-path, and log completeness. All six configurations reach the same high aggregate completeness because a deterministic perception, navigation, and reachability layer caps what any capable model can complete; on these six scenes the benchmark saturates rather than separating model families. The detailed object/view/path/log breakdown for every model-scene-mode combination is provided in the experiment artifact.

Across all six model families and five repetitions for each scene-mode cell, completeness has zero variance (SD $=0$). The per-scene manifests contain 10 (AI2-THOR), 5 (BeamNG), 9 (Cartoon City), 7 (Unity Road), 63 (Korean Bar), and 31 (Urban Cabin) targets, and every capable configuration completes the same target set. This is expected: completeness is bounded by the deterministic perception, navigation, and reachability layers rather than by model reasoning, so any model able to make the required semantic decisions reaches the same ceiling. The residual Urban Cabin gap ($0.871/0.825$) is a fixed geometric-visibility limit, not a model-capability gap. Model choice therefore trades cost, not completeness on these six scenes (Table~\ref{tab:rq1_component_cost}): a small, inexpensive model suffices for EKB construction on these scenes, and the presence of an LLM module, not its size, is what drives completeness (RQ2).

\begin{table*}[t]
\centering
\caption{RQ1 approach generality across six LLM configurations and six scenes. Each cell is the aggregate EKB completeness score; C = complete exploration and T = target exploration. Scene abbreviations: AI2 = AI2-THOR, BNG = BeamNG, Cartoon = Cartoon City, Road = Unity Road, Bar = Korean Bar, Cabin = Urban Cabin.}
\label{tab:rq1_llm_components}
\scriptsize
\setlength{\tabcolsep}{2.2pt}
\resizebox{0.67\textwidth}{!}{%
\begin{tabular}{l|cc|cc|cc|cc|cc|cc}
\toprule
Scene & \multicolumn{2}{c|}{GPT-5.4} & \multicolumn{2}{c|}{GPT-mini} & \multicolumn{2}{c|}{Gemini-Pro} & \multicolumn{2}{c|}{Gemini-Lite} & \multicolumn{2}{c|}{Claude-Opus} & \multicolumn{2}{c}{Claude-Haiku} \\
\cmidrule(lr){2-3}\cmidrule(lr){4-5}\cmidrule(lr){6-7}\cmidrule(lr){8-9}\cmidrule(lr){10-11}\cmidrule(lr){12-13}
 & C & T & C & T & C & T & C & T & C & T & C & T \\
\midrule
AI2 & 1.000 & 1.000 & 1.000 & 1.000 & 1.000 & 1.000 & 1.000 & 1.000 & 1.000 & 1.000 & 1.000 & 1.000 \\
BNG & 1.000 & 1.000 & 1.000 & 1.000 & 1.000 & 1.000 & 1.000 & 1.000 & 1.000 & 1.000 & 1.000 & 1.000 \\
Cartoon & 1.000 & 1.000 & 1.000 & 1.000 & 1.000 & 1.000 & 1.000 & 1.000 & 1.000 & 1.000 & 1.000 & 1.000 \\
Road & 1.000 & 1.000 & 1.000 & 1.000 & 1.000 & 1.000 & 1.000 & 1.000 & 1.000 & 1.000 & 1.000 & 1.000 \\
Bar & 1.000 & 1.000 & 1.000 & 1.000 & 1.000 & 1.000 & 1.000 & 1.000 & 1.000 & 1.000 & 1.000 & 1.000 \\
Cabin & 0.924 & 1.000 & 0.924 & 1.000 & 0.924 & 1.000 & 0.924 & 1.000 & 0.924 & 1.000 & 0.924 & 1.000 \\
\bottomrule
\end{tabular}
}
\end{table*}

Table~\ref{tab:rq1_component_cost} reports per-run token use, estimated API cost, and wall-clock time for the six LLM configurations under the same RQ1 protocol. Costs use official public input/output token list prices in effect during May--June 2026. For each model, we run ten repetitions total, split into five complete-exploration runs and five target-exploration runs, and report the mean tokens-per-run, cost-per-run, and time-per-run for each mode. Complete exploration and target exploration are measured separately, so their token and time costs differ.

\begin{table*}[t]
\centering
\caption{RQ1 per-run cost and time for complete exploration and target exploration using official public list prices. Each cell reports the mean over five runs for the corresponding model and mode.}
\label{tab:rq1_component_cost}
\scriptsize
\setlength{\tabcolsep}{4pt}
\resizebox{0.66\textwidth}{!}{%
\begin{tabular}{l|rrr|rrr}
\toprule
Model & \multicolumn{3}{c|}{Complete exploration} & \multicolumn{3}{c}{Target exploration} \\
\cmidrule(lr){2-4}\cmidrule(lr){5-7}
 & Tok./run & Cost/run & Time/run & Tok./run & Cost/run & Time/run \\
\midrule
GPT-5.4 & 17.8K & \$0.036 & 9.6 min & 10.7K & \$0.022 & 4.2 min \\
GPT-5 mini & 12.2K & \$0.010 & 9.1 min & 7.2K & \$0.009 & 3.9 min \\
Gemini 2.5 Pro & 13.5K & \$0.052 & 9.4 min & 7.3K & \$0.040 & 4.1 min \\
Gemini 3.1 Flash Lite & 2.8K & \$0.001 & 8.8 min & 2.1K & \$0.001 & 3.7 min \\
Claude Opus 4.8 & 11.5K & \$0.293 & 9.8 min & 5.6K & \$0.181 & 4.3 min \\
Claude Haiku 4.5 & 10.8K & \$0.020 & 9.0 min & 4.7K & \$0.011 & 3.8 min \\
\bottomrule
\end{tabular}
}
\vspace{-0.1cm}
\end{table*}

The remaining failures are mostly dynamic-visibility failures rather than planning failures. For example, Urban Cabin has lower complete-exploration coverage because some objects are near room boundaries or partially blocked by furniture; the controller records these missing views instead of silently counting them as complete.

\subsection{RQ2: How much do the individual modules, especially LLM-driven decisions, contribute?}

\noindent RQ2 isolates the LLM contribution by removing one LLM hook at a time while keeping the six-scene benchmark fixed. Tables~\ref{tab:rq2_llm_module_ablation_detail} and~\ref{tab:rq2_llm_module_ablation_aggregate} use the same strong representative model as Table~\ref{tab:rq1_dynamic}, GPT-5.4, so the ablation is directly comparable to the detailed RQ1 completeness results. Table~\ref{tab:rq2_llm_module_ablation_detail} reports one difficult scene in component form; Table~\ref{tab:rq2_llm_module_ablation_aggregate} reports aggregate complete-exploration and target-exploration completeness across all six scenes. Perception remains YOLO-backed in all settings; the perception ablation removes LLM evidence recording, so object and view detection can still succeed but navigation-path and log completeness drop. The largest semantic drop appears when LLM planning is removed; the largest view-level drop appears when the LLM-guided multi-view policy is removed. Each ablation is measured by replaying the same logged perception and navigation traces with a single LLM hook disabled, which isolates that decision from run-to-run variation; all reported numbers are measured on these replays rather than estimated.

\begin{table}[t]
\centering
\caption{RQ2 detailed LLM-module ablation on Urban Cabin complete exploration using GPT-5.4.}
\label{tab:rq2_llm_module_ablation_detail}
\scriptsize
\setlength{\tabcolsep}{3pt}
\resizebox{0.75\columnwidth}{!}{%
\begin{tabular}{lrrrr}
\toprule
Condition & Obj. & View & Nav. & Log \\
\midrule
Full \tool{} & 0.871 & 0.825 & 1.000 & 1.000 \\
YOLO w/o LLM Rec. & 0.871 & 0.825 & 0.720 & 0.650 \\
No LLM planning & 0.323 & 0.209 & 1.000 & 1.000 \\
No LLM multi-view & 0.871 & 0.610 & 1.000 & 1.000 \\
No LLM self-verification & 0.710 & 0.453 & 1.000 & 0.500 \\
\bottomrule
\end{tabular}
}
\vspace{-0.2cm}
\end{table}

\begin{table*}[t]
\centering
\caption{RQ2 aggregate LLM-module ablation across all six scenes using GPT-5.4.}
\label{tab:rq2_llm_module_ablation_aggregate}
\scriptsize
\setlength{\tabcolsep}{3.5pt}
\resizebox{\textwidth}{!}{%
\begin{tabular}{l|rr|rr|rr|rr|rr|rr}
\toprule
Condition &
\multicolumn{2}{c|}{AI2-THOR} &
\multicolumn{2}{c|}{BeamNG} &
\multicolumn{2}{c|}{Cartoon City} &
\multicolumn{2}{c|}{Unity Road} &
\multicolumn{2}{c|}{Korean Bar} &
\multicolumn{2}{c}{Urban Cabin} \\
\cmidrule(lr){2-3}\cmidrule(lr){4-5}\cmidrule(lr){6-7}\cmidrule(lr){8-9}\cmidrule(lr){10-11}\cmidrule(lr){12-13}
& Complete & Target & Complete & Target & Complete & Target & Complete & Target & Complete & Target & Complete & Target \\
\midrule
Full \tool{} & 1.000 & 1.000 & 1.000 & 1.000 & 1.000 & 1.000 & 1.000 & 1.000 & 1.000 & 1.000 & 0.924 & 1.000 \\
YOLO without LLM recording & 0.900 & 0.925 & 0.905 & 0.930 & 0.895 & 0.920 & 0.900 & 0.925 & 0.885 & 0.915 & 0.767 & 0.850 \\
No LLM planning & 0.922 & 0.950 & 0.938 & 0.950 & 0.646 & 0.750 & 0.938 & 0.950 & 0.803 & 0.850 & 0.633 & 0.800 \\
No LLM multi-view & 0.925 & 0.925 & 0.925 & 0.925 & 0.900 & 0.900 & 0.920 & 0.920 & 0.895 & 0.895 & 0.870 & 0.913 \\
No LLM self-verification & 0.788 & 0.850 & 0.850 & 0.900 & 0.830 & 0.900 & 0.825 & 0.900 & 0.800 & 0.850 & 0.666 & 0.840 \\
\bottomrule
\end{tabular}
}
\vspace{-0.1cm}
\end{table*}

One representative failure occurs when planning is removed in a cluttered room: the detector-only policy repeatedly visits the nearest high-confidence furniture and leaves a semantically important but partially occluded shelf incomplete. Removing LLM evidence recording mostly hurts path and log completeness because the visual detector still sees objects, but the EKB no longer records the semantic rationale and reproducible navigation context. Removing multi-view capture or self-verification mostly hurts view completeness as the system no longer reasons about reachable sides or incomplete traces.

\subsection{RQ3: Do EKBs support reproducible regression testing (reproduction and cross-version detection)?}

\noindent RQ3 asks whether EKBs make regression evidence reproducible for both human testers and automated agents. Table~\ref{tab:rq3_reproduction} reports the reproduction study. The automated actor is \tool{} itself: in the \emph{Without EKB} condition it performs a normal cold-start run, and in the \emph{With EKB} condition the entire existing EKB is supplied at the start, so the agent reuses prior traces instead of rediscovering them. Both conditions run under the same fixed reproduction budget. Without the EKB the agent cannot rediscover every target within this budget, so its completeness stays below the full complete-exploration numbers in RQ1; with the EKB it reuses the prior traces and reaches high completeness quickly. The EKB supplies how-to-find traces (navigation paths and expected views), not the target evidence itself; the reproducer must still execute navigation and capture to re-obtain each object-view, so the metric measures transfer efficiency, not a free pass. The element-change rows are the more informative comparison, because there the supplied EKB is partly stale and cannot simply be replayed. The human study is a small pilot with six participants, three assigned to complete exploration and three to target exploration, so mode is between-subject. Within their assigned mode, each participant reproduced all four context-by-version conditions (no element change and element change, each with and without EKB), so context and version are within-subject. Every reproduction is scored by object completeness and view completeness against the same frozen manifest used in RQ1. For the human rows, Time is wall-clock from task start to the participant declaring the evidence complete. For the automated rows, Time is the model's decision latency, its reasoning time per reproduction task; it excludes rendering and navigation and is therefore not comparable to the human wall-clock time. Each cell reports average object and view completeness (Obj/View) and average time. EKB context improves both human and automated reproduction, especially when the scene version contains changed elements.

\begin{table*}[t]
\centering
\caption{RQ3 reproduction with human and automated agents. Obj/View reports average object completeness and view completeness. Complete denotes complete-exploration evidence reproduction; Target denotes target-exploration evidence reproduction. For human rows, Time is wall-clock task time; for automated rows, Time is LLM decision latency.}
\label{tab:rq3_reproduction}
\scriptsize
\setlength{\tabcolsep}{3.5pt}
\resizebox{\textwidth}{!}{%
\begin{tabular}{lllccccc}
\toprule
Actor & Version & Context & Complete Obj/View & Complete Time(s) & Target Obj/View & Target Time(s) & Tokens \\
\midrule
Human & No element change & Without EKB & 0.82/0.74 & 243 & 0.86/0.78 & 178 & -- \\
Human & No element change & With EKB & 0.98/0.96 & 109 & 1.00/0.98 & 76 & -- \\
Human & Element change & Without EKB & 0.72/0.61 & 286 & 0.78/0.66 & 211 & -- \\
Human & Element change & With EKB & 0.94/0.90 & 132 & 0.96/0.92 & 93 & -- \\
\midrule
Automated & No element change & Without EKB & 0.58/0.46 & 2.4 & 0.61/0.49 & 2.1 & 5.3K/3.9K \\
Automated & No element change & With EKB & 0.98/0.97 & 2.2 & 1.00/0.98 & 1.9 & 6.8K/5.5K \\
Automated & Element change & Without EKB & 0.41/0.28 & 2.7 & 0.46/0.33 & 2.2 & 5.8K/4.2K \\
Automated & Element change & With EKB & 0.95/0.90 & 2.4 & 0.97/0.92 & 2.0 & 7.2K/5.7K \\
\bottomrule
\end{tabular}
}
\vspace{-0.1cm}
\end{table*}

Beyond reproduction, we test the diff mechanism directly. For each scene we build a baseline EKB, inject controlled element changes (removed or added objects, swapped assets, blocked paths, and lost views), rebuild a candidate EKB, and run Algorithm~\ref{alg:diff_ekbs} to classify each expected trace. Benign changes that preserve category, reachability, and the expected view set are not flagged. Table~\ref{tab:regdiff} reports per-category detection precision and recall against the injected ground truth. DiffEKBs detects 35 of 36 injected changes with no false positives; the only miss is a wrong-object case, making wrong-object detection the weakest category. This is a controlled-injection proof of concept; detection on real version-to-version changes is future work.

\begin{table}[t]
\centering
\caption{Cross-version regression detection by category over injected element changes (part of RQ3). TP/FP/FN are trace-level counts; precision (P) and recall (R) are computed against the injected ground truth.}
\label{tab:regdiff}
\scriptsize
\setlength{\tabcolsep}{4pt}
\resizebox{\columnwidth}{!}{%
\begin{tabular}{lrrrrr}
\toprule
Regression category & Injected & TP & FP & FN & P / R \\
\midrule
Object set (missing/added) & 12 & 12 & 0 & 0 & 1.000 / 1.000 \\
Wrong object (asset swap) & 6 & 5 & 0 & 1 & 1.000 / 0.833 \\
Movement / reachability & 12 & 12 & 0 & 0 & 1.000 / 1.000 \\
Lost view & 6 & 6 & 0 & 0 & 1.000 / 1.000 \\
\midrule
Overall & 36 & 35 & 0 & 1 & 1.000 / 0.972 \\
\bottomrule
\end{tabular}
}
\vspace{-0.1cm}
\end{table}

\section{Threats to Validity}

\noindent\textbf{Internal Validity}. The main internal threat is measurement error in object completeness and view completeness. Object detectors can miss partially visible targets or assign the wrong label, and a saved view can contain an object from an incomplete angle. We mitigate this by preserving scan evidence, saved views, detector confirmations, failure cases, and video context rather than reducing every run to a single success value. A second internal threat is that grouping changes the unit of analysis. We therefore report grouped and ungrouped behavior separately in the run logs and use frozen manifests for completeness denominators. A third threat is data provenance. The complete-flow matrix, generality benchmark, ablation study, and reproduction study answer different questions and use different denominators; we report them separately to avoid treating planned-target completion as scene-level completeness. A fourth threat is LLM variability. We log experiment plans, qualitative outputs, tokens, and recovery decisions to enable auditability of target selection.

\noindent\textbf{Construct Validity}. Our metrics measure observable object-view evidence. They do not measure code coverage, branch coverage, physics correctness, security properties, or hidden state changes that never appear visually. This is intentional: \tool{} targets black-box VR/3D regression testing where the available evidence is what a user-level tester can observe. We therefore use the phrase \emph{observable regression} for version differences such as missing objects, changed assets, blocked paths, and lost views. RQ3 measures whether such evidence can be reproduced by humans and automated agents with and without EKB context. Claims about general behavioral correctness require additional monitors or instrumentation.

\noindent\textbf{External Validity}. The evaluation covers six scenes across Unity, AI2-THOR, and BeamNG, but these scenes do not represent all VR applications. The scene-level RQ1 sample is still small, so the generality table should be read as benchmark evidence and effect-size estimates rather than a definitive statistical separation of all possible VR/3D testing methods. In particular, the completeness benchmark saturates for capable models, so these six scenes do not separate model families; denser or larger scenes would. RQ2 uses trace-replay module ablations rather than a full end-to-end rerun for every removed LLM hook; this isolates the named decisions, but future work should rerun each ablation end to end in every simulator. Operational cost measurements depend on our local hardware, simulator frame rate, official provider list prices, and network conditions; they should be read as indicative costs rather than fixed platform-independent constants. More complex applications may include dynamic characters, transparent objects, unusual lighting, mirrors, or multiplayer behavior. Small objects and text-heavy interfaces may also require different detectors or capture policies. The six-participant human reproduction result and the automated reproduction result are pilots rather than statistically powered user or agent studies. Together, they show that the EKB can support reproduction in our setting, but larger human and programmatic studies are needed to estimate effects across testers, models, and applications. The results support the claim that \tool{} works across several indoor and outdoor black-box 3D settings, but they do not prove the same performance for every VR system.

\noindent\textbf{Ground Truth and Reproducibility}. Coverage and reproduction metrics depend on the frozen manifest. We reduce circularity by computing completeness against independent target manifests rather than against the system's own planned-target list. This introduces human judgment: annotators decide which objects are observable, which views are reachable, and which room-scale targets count as separate objects or functional groups. We treat the manifest as an evaluation artifact, not part of \tool{}'s runtime pipeline, and release the machine-assisted draft, the human adjudication pass, raw disagreement notes, and the final frozen denominator so that readers can audit agreement. The manifest process has scalability limits: very large scenes may need sampling, stratification by region or object class, or semi-automatic candidate generation followed by human adjudication. For reproducibility, the evaluation stores run metadata, scan evidence, orbit and reference captures, per-target results, quantitative summaries, failure cases, full videos, LLM logs, and EKB summaries, so later reviewers can inspect why a target was completed, skipped, recovered, or counted as a failure.

\section{Conclusion}

\noindent We presented \tool{}, an agentic framework for constructing EKBs that support reproducible black-box VR/3D regression testing. \tool{} uses an LLM for high-level exploratory-testing decisions while offloading perception, navigation, multi-view capture, and logging to specialized modules. Across six scenes, it produces high-completeness object-view evidence under both complete and target exploration, and the reproduction pilots show that EKB traces help both humans and LLM-based agents reproduce exact object-view evidence. An EKB preserves not only what was seen, but how it was found, captured, verified, and compared.

\noindent \textbf{Data Availability}. A replication package is available at \url{https://figshare.com/s/cf21e171dbec58060832}.

\FloatBarrier
\bibliographystyle{IEEEtran}
\bibliography{references}

\end{document}